\documentclass[twocolumn,aps,showpacs,prl]{revtex4}
\usepackage{graphicx}
\usepackage{epstopdf}
\usepackage{amsmath}
\usepackage{multirow}

\begin{document}

\title{Ternary iron selenide K$_{0.8}$Fe$_{1.6}$Se$_2$
is an antiferromagnetic semiconductor}

\author{Xun-Wang Yan$^{1,2}$}
\author{Miao Gao$^{1}$}
\author{Zhong-Yi Lu$^{1}$}\email{zlu@ruc.edu.cn}
\author{Tao Xiang$^{2,3}$}\email{txiang@iphy.ac.cn}

\date{\today}

\affiliation{$^{1}$Department of Physics, Renmin University of
China, Beijing 100872, China}

\affiliation{$^{2}$Institute of Theoretical Physics, Chinese Academy
of Sciences, Beijing 100190, China }

\affiliation{$^{3}$Institute of Physics, Chinese Academy of
Sciences, Beijing 100190, China }

\begin{abstract}

We have studied electronic and magnetic structures of
K$_{0.8+x}$Fe$_{1.6}$Se$_2$ by performing the first-principles
electronic structure calculations. The ground state of the
Fe-vacancies ordered K$_{0.8}$Fe$_{1.6}$Se$_2$ is found to be a
quasi-two-dimensional blocked checkerboard antiferromagnetic (AFM)
semiconductor with an energy gap of 594 meV and a large ordering
magnetic moment of 3.37 $\mu_B$ for each Fe atom, in excellent
agreement with the neutron scattering measurement. The underlying
mechanism is the chemical-bonding-driven tetramer lattice
distortion. K$_{0.8+x}$Fe$_{1.6}$Se$_2$ with finite $x$ is a doped
AFM semiconductor with low conducting carrier concentration which is
approximately proportional to the excess potassium content,
consistent qualitatively with the infrared observation. Our study
reveals the importance of the interplay between antiferromagnetism
and superconductivity in these materials. This suggests that
K$_{0.8}$Fe$_{1.6}$Se$_2$, instead of KFe$_2$Se$_2$, should be
regarded as a parent compound from which the superconductivity
emerges upon electron or hole doping.

\end{abstract}

\pacs{74.70.Xa, 74.20.Pq, 74.20.Mn}

\maketitle


The recent discovery of high-Tc superconductivity in potassium
intercalated FeSe\cite{chen} and other iron-based
chalcogenides\cite{Cs,fang} has triggered a surge of interest for
the investigation of unconventional superconducting pairing
mechanism. In particular, it was found that the superconductivity in
these materials coexists with a strong antiferromagnetic (AFM) order
with an unprecedentedly large magnetic moment of 3.31 $\mu_B$/Fe
formed below a Neel temperature of 559$K$ \cite{muSR,bao1},
meanwhile the conducting electron concentration is extremely
low\cite{infrared}. Unlike the collinear \cite{cruz,ma1,yan} or
bi-collinear\cite{ma,bao,shi} AFM order observed in the parent
compounds of other iron-based superconductors, the neutron
observation found that these materials have a blocked checkerboard
AFM order\cite{bao1}.

These compounds have the ThCr$_{2}$Si$_{2}$ type crystal structure,
as shown in Fig. 1(a), isostructural with 122-type iron pnictides
\cite{rotter}. From the latest X-rays, transmission electron
microscopy, and neutron scattering measurements, it was suggested
that the composition of the K-intercalated FeSe superconductors
K$_y$Fe$_{2-x}$Se$_2$ is close to K$_y$Fe$_{1.6}$Se$_2$ with a
fivefold expansion of the parent ThCr$_2$Si$_2$ unit cell in the
$ab$ plane, namely a $\sqrt{5}\times \sqrt{5}$ Fe-vacancies
superstructure \cite{basca,bao1}, which corresponds to one-fifth
Fe-vacancies. In an Fe-Fe square lattice, an Fe atom nearby a
vacancy is 3-Fe-neighbored, or 2-Fe-neighbored, or even
1-Fe-neighbored. Topologically a square lattice composed only by
3-Fe-neighbored Fe atoms exists only in one arranging order but with
two different kinds of chirality, right-handed or left-handed
rotations\cite{sab}, as shown in Fig. 1(b) and (c), respectively.
This is exactly a $\sqrt{5}\times \sqrt{5}$ superstructure.

The peculiar properties observed in K-intercalated iron-based
chalcogenide superconductors indicate that these materials represent
a special limit where the interplay between antiferromagnetism and
superconductivity plays an important role in the formation of
superconducting pairs. To reveal the hidden physics, we have
performed the first-principles electronic structure calculations for
K$_{0.8+x}$Fe$_{1.6}$Se$_2$ with $x=0$ and $\pm0.1$. We find that
the ground state of K$_{0.8}$Fe$_{1.6}$Se$_2$ is an AFM
semiconductor with a blocked checkerboard long-range AFM order (Fig.
3(c)) and an energy gap of 594 meV, and K$_{0.8\pm
0.1}$Fe$_{1.6}$Se$_2$ are doped electron- or hole-type AFM
semiconductors. K$_{0.8}$Fe$_{1.6}$Se$_2$ is thus an essential
parent compound so that the observed superconductivity in
K$_y$Fe$_{2-x}$Se$_2$ is based on the doped AFM semiconductors.

\begin{figure}[t]
\includegraphics[width=8.0cm]{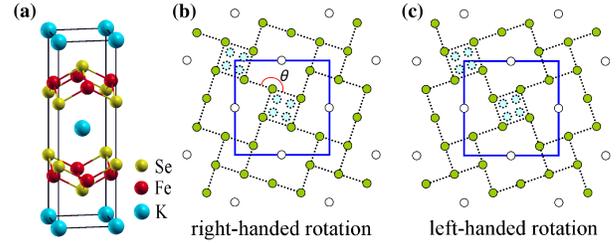}
\caption{(Color online) K$_y$Fe$_x$Se$_2$ with the
ThCr$_{2}$Si$_{2}$-type structure: (a) a tetragonal unit cell
containing two formula units without any vacancy; (b) and (c) are
schematic top view of an Fe-Fe square layer with one-fifth
Fe-vacancies ($x=1.6$) ordered with the right- and left-chirality,
respectively. In this $\sqrt{5}\times \sqrt{5}$ lattice, each Fe is
coordinated with 3 neighboring Fe atoms. The squares enclosed by the
solid lines denote the unit cells. The filled circles denote the Fe
atoms while the empty circles denote the Fe-vacancies. After
optimizing the structure by the energy minimization, the Fe atoms go
into a blocked distribution from a uniform square distribution, in
which the four closest Fe atoms form a block, represented by dashed
(blue) circles in (b) and (c), whose bond distance is shorter than
that of the uniform Fe square lattice (green dots). Correspondingly,
the angle $\theta$ decreases from 180$^{\circ}$ to $176^{\circ}-
178^{\circ}$. } \label{figa}
\end{figure}

In our calculations the plane wave basis method was used. We adopted
the generalized gradient approximation (GGA) with
Perdew-Burke-Ernzerhof formula \cite{pbe} for the
exchange-correlation potentials. The ultrasoft pseudopotentials
\cite{vanderbilt} were used to model the electron-ion interactions.
The kinetic energy cut-off and the charge density cut-off of the
plane wave basis were chosen to be 800 eV and 6400 eV, respectively.
A mesh of $18\times 18\times 8$ k-points were sampled for the
Brillouin-zone integration while the Gaussian broadening technique
was used in the case of metallic states. In the calculations, the
lattice parameters with the internal atomic coordinates were
optimized by the energy minimization. The optimized tetragonal
lattice parameters in the magnetic states agree excellently with the
experimental data\cite{bao1}.

In the calculations, we adopted a $\sqrt{5}\times\sqrt{5}\times 1$
tetragonal supercell, in which there are two FeSe layers with total
16 Fe atoms and 4 Fe-vacancies, 20 Se atoms, and 8 K atoms and 2
K-vacancies, to represent K$_{0.8}$Fe$_{1.6}$Se$_2$. For K atoms,
there are two inequivalent positions in the supercell. Accordingly,
there are a number of inequivalent arrangements for the two
K-vacancies. From the calculations, we find that both the electronic
band structures and the Fe moments are hardly affected by these
inequivalent arrangements. And the energy difference among these
different K arrangements is within 1 meV/Fe in both spin-polarized
and spin-unpolarized cases. This is consistent with the experimental
observation that the vacancies on K sites are in positional disorder
\cite{basca}.

\begin{figure}[tb]
\includegraphics[width=6.0cm]{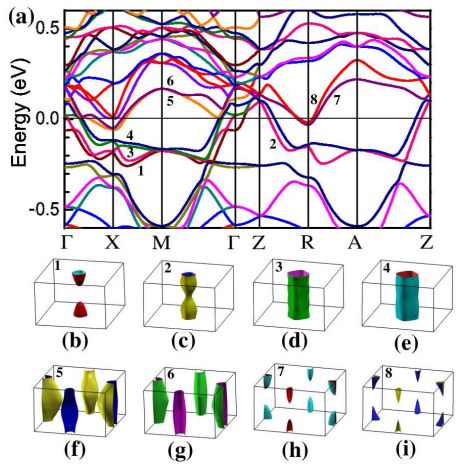}
\caption{(Color online) Electronic structure of
K$_{0.8}$Fe$_{1.6}$Se$_2$ with one-fifth Fe-vacancies ordered in a
way as shown by Fig. 1(b) or (c) in the nonmagnetic state: (a) the
band structure; (b-i) are the Fermi surface sheets due to the bands
crossing the Fermi energy denoted by (1-8) in (a), respectively. The
corresponding Brillouin zone is shown in Fig. 4(b). The Fermi energy
sets to zero.}
\end{figure}

Figure 2 shows the non-magnetic band structure and Fermi surface of
K$_{0.8}$Fe$_{1.6}$Se$_2$ with one-fifth Fe-vacancies ordered in a
way as shown by Fig. 1(b) or (c). There are 8 bands crossing the
Fermi level. An important feature revealed by the calculation is
that due to a substantial lattice distortion driven by Fe vacancies,
the bond distances among four closest Fe atoms around a Se atom are
reduced. These four Fe atoms will form a compact square (we will
call it a block below), as shown by the light blue circles in Fig.
1(b). The angle $\theta$, which is shown in Fig. 1(b), decreases
from 180$^{\circ}$ to about 176$^{\circ}$. As a result, the Fe-Fe
and Fe-Se chemical bonds within each block become stronger. This
gains an energy of about 160 meV/Fe. Such a tetramer lattice
distortion is not due to the spin-phonon interaction since it
happens in both non-magnetic and magnetic states. It is in fact
driven by the chemical bonding, namely this is a
chemical-bonding-driven lattice distortion. On the contrary, all the
structural or lattice distortions previously found in the
iron-pnictides and chalcogenides are driven by the spin-phonon
interactions \cite{ma1}, namely magnetism-driven lattice
distortions.

\begin{figure}[tb]
\includegraphics[width=6.5cm]{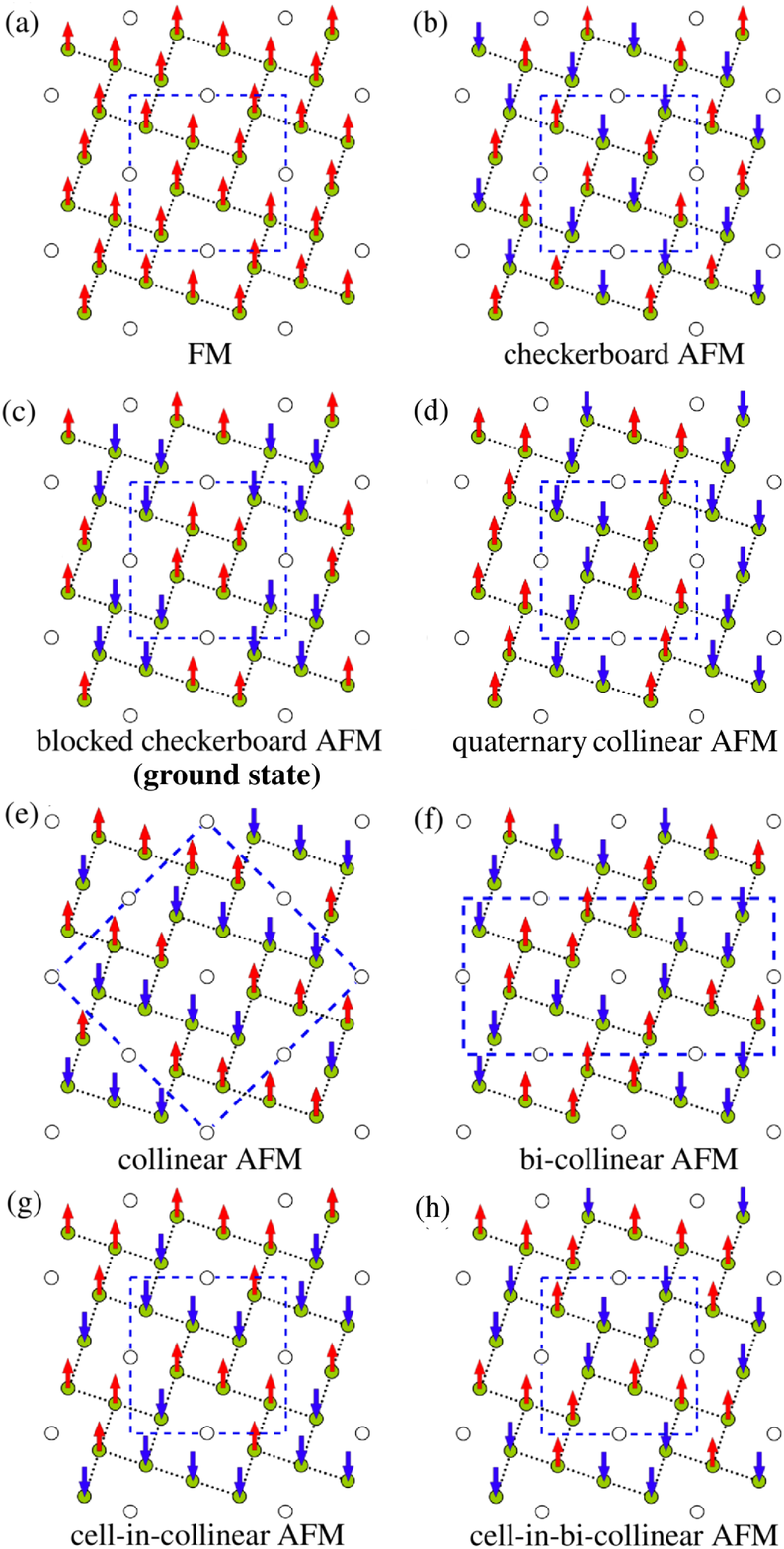}
\caption{(Color online) Schematic top view of eight possible
magnetic orders in the Fe-Fe square layer with one-fifth
Fe-vacancies. Among these magnetic orders, the blocked checkerboard
antiferromagnetic order in (c) is the ground state, in which the
magnetic moments within each block are in parallel, but the magnetic
moments for the neighboring two blocks are in anti-parallel.  The Fe
spins are shown by arrows. The empty circles mark the Fe vacancies.
The squares or rectangles enclosed by the dashed lines denote the
magnetic unit cells.} \label{fig3}
\end{figure}

Figure 3 shows eight possibly energetically favored magnetic orders
arranged in a right-chirality lattice (Fig. 1(b)). From the
calculations, if the energy of the nonmagnetic state is set to zero,
we find that the respective energies of these eight states, i.e. the
ferromagnetic, checkerboard AFM, blocked checkerboard AFM,
quaternary collinear AFM, collinear AFM, bi-collinear AFM, cell-in
collinear AFM, and cell-in bi-collinear AFM states, are (-0.070,
-0.189, -0.342, -0.232, -0.293, -0.229, -0.254, -0.267) eV/Fe for
K$_{0.8}$Fe$_{1.6}$Se$_2$. The corresponding magnetic moments are
(3.29, 2.76, 3.37, 3.10, 3.32, 3.22, 3.20, 3.26) $\mu_B$/Fe. The
ground state of K$_{0.8}$Fe$_{1.6}$Se$_2$ is thus found to have a
blocked checkerboard AFM order (Fig. 3(c)) with a large magnetic
moment of 3.37$\mu_B$/Fe, in agreement excellently with the latest
neutron scattering measurement \cite{bao1}. The similar magnetic
order was also reported in a calculation to show metallic
KFe$_{1.6}$Se$_2$ \cite{cao}.

Here we emphasize that it is the chemical-bonding driven tetramer
lattice distortion mentioned above that makes the blocked
checkerboard AFM order be the lowest in energy. Without this
tetramer lattice distortion, the energy of the collinear AFM state
(Fig. 3(e)) is in fact 43 meV/Fe lower than that of the blocked
checkerboard AFM state. The collinear AFM order is the ground state
magnetic order of the iron-pnictides, driven by the As-bridged AFM
superexchange interaction between two next-nearest Fe atoms, which
dominates over other exchange interactions. However, in
K$_{0.8}$Fe$_{1.6}$Se$_2$, the chemical-bonding driven tetramer
lattice distortion reduces the bond distance within each block so
that the ferromagnetic exchange interaction between the two nearest
Fe atoms inside a block and the AFM exchange interactions between
neighboring two blocks are substantially enhanced. It turns out that
the blocked checkerboard AFM order dominates over all the other
magnetic orders.

\begin{figure}
\includegraphics[width=7.5cm]{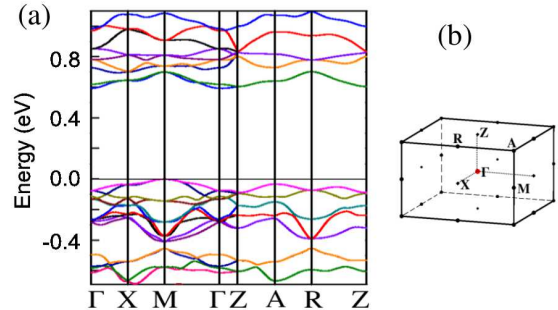}
\caption{(Color online) (a) Electronic band structure of
Fe-vacancies ordered K$_{0.8}$Fe$_{1.6}$Se$_2$ in the ground state
with a blocked checkerboard antiferromagnetic order (Fig. 3(c)). (b)
Brillouin zone. Here the top of the valence band is set to zero. }
\end{figure}

Figure 4(a) shows the electronic band structure of
K$_{0.8}$Fe$_{1.6}$Se$_2$ in the blocked checkerboard AFM ground
state. To our surprise, we find that the compound
K$_{0.8}$Fe$_{1.6}$Se$_2$ is an AFM semiconductor with an energy
band gap as large as $\sim 594$ meV. This is different from the
parent compounds of other iron-based superconductors, they are
either magnetic or non-magnetic semimetals.
K$_{0.8}$Fe$_{1.6}$Se$_2$ with the checkerboard AFM, collinear AFM,
or quaternary collinear AFM order is also found to be
semiconducting. The corresponding energy band gaps are 22, 73, and
185 meV, respectively. With any other magnetic order,
K$_{0.8}$Fe$_{1.6}$Se$_2$ is in a metallic state.

\begin{figure}
\includegraphics[width=6.5cm]{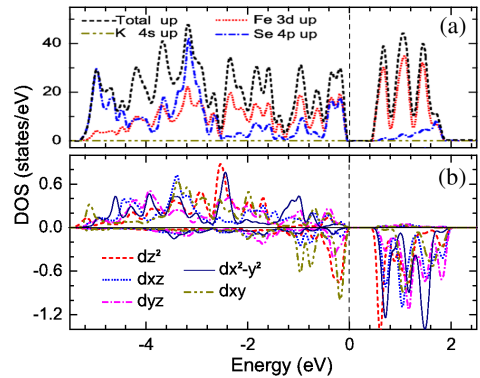}
\caption{(Color online) K$_{0.8}$Fe$_{1.6}$Se$_2$ in the blocked
checkerboard antiferromagnetic ground state: (a) total and
orbital-resolved partial density of states (spin-up part); (b)
projected density of states onto the five Fe-$3d$ orbitals at an Fe
atom. Here the top of the valence band sets to zero. And $x$-axis is
along the two next-nearest neighbor Fe-Fe.}
\end{figure}

Figure 5(a) shows the total and orbital-resolved partial density of
states (spin-up part) for K$_{0.8}$Fe$_{1.6}$Se$_2$ in the blocked
checkerboard AFM ground state. We find that the states around the
top valence bands are composed of both Fe-3d and Se-4p orbitals with
the same peak energies. This is consistent with the localized
features of these bands, enhanced by Fe-Fe and Fe-Se chemical
bonding due to the tetramer lattice distortion. On the contrary, the
lower conduction bands are mainly contributed by the Fe-3d orbitals.

By projecting the density of states onto the five $3d$ orbitals of
an Fe atom in the blocked checkerboard AFM ground state (Fig. 5(b)),
we find that the five up-spin orbitals are almost fully filled. This
suggests that the crystal field splitting induced by Se atoms is
small, similar to the case of iron-pnictides \cite{ma1}. The large
magnetic moment formed around each Fe results from the Hund's rule
coupling. The down spin orbitals are partially filled by $d_{z^2}$,
$d_{xy}$, and $d_{yz}$ orbitals. Such anisotropy among the five down
spin orbitals in K$_{0.8}$Fe$_{1.6}$Se$_2$ is enhanced by the
ordered Fe-vacancies.

We have verified the above results using the virtual crystal
approximation (VCA) for K atoms. We then replaced K$_{0.8}$ by
K$_{0.7}$ and K$_{0.9}$ respectively by using VCA to simulate the
hole-doping and electron-doping effect upon the AFM semiconductor
K$_{0.8}$Fe$_{1.6}$Se$_2$. As shown in Fig. 6, the electronic band
structure of hole or electron doped K$_{0.8\pm 0.1}$Fe$_{1.6}$Se$_2$
is almost the same as the one for the un-doped case (Fig. 4). The
magnetic moments of Fe ions are also hardly changed by doping. The
doped electrons and holes occupy the bottom conduction bands and the
top valence bands, respectively. The volumes enclosed by the Fermi
surfaces are found to be 0.529 holes/cell and 0.642 electrons/cell,
namely $5.00\times 10^{20}$holes/$cm^3$ and $6.06\times
10^{20}$electrons/$cm^3$, respectively. In principle, every excess
K$_{0.1}$ can supply 1 electron/cell to the material. Thus both
electron and hole doping upon K$_{0.8}$Fe$_{1.6}$Se$_2$ are very
efficient, similar to conventional semiconductors. But the total
carrier concentration is very low, in comparison with other
iron-based superconductors. This agrees with the infrared
measurement\cite{infrared}.

We have calculated the formation energy of
K$_{0.8+x}$Fe$_{1.6-x/2}$Se$_2$, defined by  $(0.8+x)E_K+(1.6-x/2)
E_{Fe}+2E_{Se} - E_x$, where $E_K$, $E_{Fe}$, and $E_{Se}$ are the
atomic energies of K, Fe, and Se, respectively, and $E_x$ is the
total energy of K$_{0.8+x}$Fe$_{1.6-x/2}$Se$_2$. For $x=0$, 0.2 and
$-0.4$, corresponding to K$_{0.8}$Fe$_{1.6}$Se$_2$,
KFe$_{1.5}$Se$_2$, and K$_{0.4}$Fe$_{1.8}$Se$_2$, we find that the
formation energies are 16.652, 16.564, and 16.632 eV, respectively.
Here these stoichiometric compounds are in a balance of chemical
valence, in which K, Fe and Se are in 1+, 2+, and 2- valence states,
respectively. The ground states of Fe-vacancies ordered
K$_{0.4}$Fe$_{1.8}$Se$_2$ and KFe$_{1.5}$Se$_2$ are in a collinear
AFM order\cite{yan}. The formation energy data indicate that
K$_{0.8}$Fe$_{1.6}$Se$_2$ is energetically more favorable when K
atoms are intercalated into FeSe. Thus K$_{0.8}$Fe$_{1.6}$Se$_2$ is
more likely to be the parent compound of potassium intercalated FeSe
superconductors upon doping. The superstructure $\sqrt{5}\times
\sqrt{5}$ with Fe$_{1.6}$Se$_2$ layers may thus be the framework or
backbone of superconductors K$_y$Fe$_{2-x}$Se$_2$.

\begin{figure}[tb]
\includegraphics[width=7.0cm]{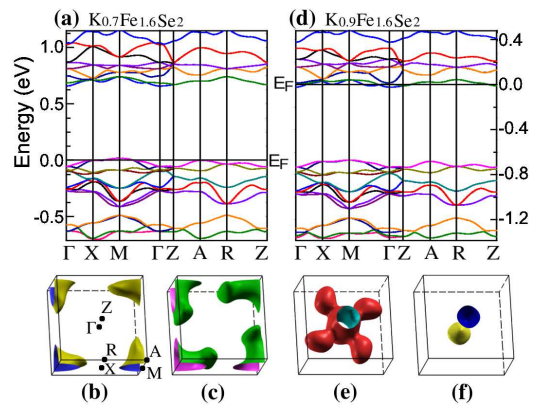}
\caption{(Color online) Electronic band structures and Fermi
surfaces in the blocked checkerboard antiferromagnetic ground state
with the left chirality: (a-c) for K$_{0.7}$Fe$_{1.6}$Se$_2$ and
(d-f) for K$_{0.9}$Fe$_{1.6}$Se$_2$, to represent the hole-doping or
electron-doping upon K$_{0.8}$Fe$_{1.6}$Se$_2$, respectively. The
Fermi energy sets to zero.}
\end{figure}

We have also calculated the electronic structures of
A$_{0.8}$Fe$_{1.6}$Se$_2$ (A=Rb, Cs, or Tl) and found that they are
all AFM semiconductors. The energy band gaps of these compounds are
571, 548, and 440 meV, respectively.

In conclusion, we have performed the first principles calculations
for the electronic structure and magnetic order of
K$_y$Fe$_{2-x}$Se$_2$. We find that the ground state of
K$_{0.8}$Fe$_{1.6}$Se$_2$ is a quasi-two-dimensional
antiferromagnetic semiconductor with an energy gap of 594 meV. The
ground state magnetic order is a blocked checkerboard
antiferromagnetic order with a large magnetic moment of 3.37$\mu_B$,
in agreement with the experimental measurements. The underlying
mechanism is the chemical-bonding-driven tetramer lattice
distortion. Doping electrons or holes into K$_{0.8}$Fe$_{1.6}$Se$_2$
by intercalating more or less K to the compound almost do not change
the band structure as well as the blocked checkerboard
antiferromagnetic order. Our result suggests that
K$_{0.8}$Fe$_{1.6}$Se$_2$ serves as a parent compound which becomes
superconducting upon electron or hole doping.

We would like to thank W. Bao for helpful discussion and P.C. Dai
for drawing our attention for the magnetic ordered state as shown in
Fig. 3(d). This work is partially supported by National Natural
Science Foundation of China and by National Program for Basic
Research of MOST, China.

\end{document}